\PassOptionsToPackage{table,xcdraw}{xcolor}
\documentclass[conference]{IEEEtran}
\IEEEoverridecommandlockouts
\usepackage{xcolor}
\usepackage{cite}
\usepackage{amsmath,amssymb,amsfonts}
\usepackage{gensymb}
\usepackage{algorithmic}
\usepackage{graphicx}
\usepackage{subcaption}
\usepackage{nicematrix}
\usepackage{physics}
\usepackage{textcomp}
\usepackage{multirow}
\usepackage{booktabs}
\usepackage{float}
\usepackage{hyperref}
\def\BibTeX{{\rm B\kern-.05em{\sc i\kern-.025em b}\kern-.08em
    T\kern-.1667em\lower.7ex\hbox{E}\kern-.125emX}}
\begin{document}

\title{Comparative Analysis of QNN Architectures for Wind Power Prediction: Feature Maps and Ansatz Configurations}

\author{\IEEEauthorblockN{Batuhan Hangun}
\IEEEauthorblockA{\textit{Computer Engineering Department} \\
\textit{Yildiz Technical University}\\
İstanbul, Türkiye \\
batuhanhangun@gmail.com}
\and
\IEEEauthorblockN{Emine Akpinar}
\IEEEauthorblockA{\textit{Department of Physics} \\
\textit{Yildiz Technical University}\\
İstanbul, Türkiye \\
emineakpinar28@gmail.com}
\and
\IEEEauthorblockN{Oguz Altun}
\IEEEauthorblockA{\textit{Computer Engineering Department} \\
\textit{Yildiz Technical University}\\
İstanbul, Türkiye \\
oaltun@yildiz.edu.tr}
\and
\IEEEauthorblockN{Onder Eyecioglu}
\IEEEauthorblockA{\textit{Computer Engineering Department} \\
\textit{Bolu Abant Izzet Baysal University}\\
Bolu, Türkiye \\
onder.eyecioglu@ibu.edu.tr}
}

\maketitle

\begin{abstract}
Quantum Machine Learning (QML) is an emerging field at the intersection of quantum computing and machine learning, aiming to enhance classical machine learning methods by leveraging quantum mechanics principles such as entanglement and superposition. However, skepticism persists regarding the practical advantages of QML, mainly due to the current limitations of noisy intermediate-scale quantum (NISQ) devices. This study addresses these concerns by extensively assessing Quantum Neural Networks (QNNs)—quantum-inspired counterparts of Artificial Neural Networks (ANNs), demonstrating their effectiveness compared to classical methods. We systematically construct and evaluate twelve distinct QNN configurations, utilizing two unique quantum feature maps combined with six different entanglement strategies for ansatz design. Experiments conducted on a wind energy dataset reveal that QNNs employing the Z feature map achieve up to 93\% prediction accuracy when forecasting wind power output using only four input parameters. Our findings show that QNNs outperform classical methods in predictive tasks, underscoring the potential of QML in real-world applications.
 
\end{abstract}

\begin{IEEEkeywords}
Quantum machine learning, quantum neural networks, quantum computing, wind power prediction, renewable energy
\end{IEEEkeywords}

\section{Introduction}

Quantum computing is still in its infancy but is rapidly maturing through ongoing global research efforts. The limitations imposed by the Noisy Intermediate-Scale Quantum (NISQ) era contribute to skepticism regarding the real-world applicability of quantum computing. To address these concerns, researchers must demonstrate that Quantum Machine Learning (QML) performs competitively against classical methods.

Currently, accelerating artificial intelligence (AI) research is considered one of the most promising applications of quantum computing. Consequently, QML focuses on developing quantum algorithms analogous to classical machine learning (ML) methods, such as support vector machines (SVM), to enhance performance in solving real-world problems.

In this study, we conducted an experimental approach to investigate various Quantum Neural Network (QNNs) structures and their performance on a prediction problem. We used two different feature maps and combined them with six different ansatz circuits. We compared the QNN configurations with established classical approaches. Our experiments demonstrate that the feature map and ansatz-based QNN implementations can effectively address machine learning tasks. Furthermore, they exhibit superior performance compared to the classical approaches. These promising results indicate that QNN-based solutions utilizing the Z feature map are valuable implementations for advancing the field of QML for prediction tasks.

\section{Related Works}
Research across various fields indicates significant potential for QML applications. Kyriienko and Magnusson \cite{kyriienko2022} proposed an unsupervised QML approach for credit card fraud detection. Their application of quantum kernels to fraud detection resulted in a 15\% increase in precision compared to classical kernels \cite{kyriienko2022}. In the study of Bhowmik and Thapliyal, variational quantum circuits (VQC)-based QML approach was introduced to detect anomalies that are common in consumer electronics \cite{Bhowmik2024}. In another study, Cultice et al. deployed a quantum-hybrid Support Vector Machine (SVM) for solving anomaly detection problems in the cyber-physical systems domain. They demonstrated that a QML approach can yield up to a 14\% performance advantage \cite{Cultice2024}. Early applications of QNN demonstrated superior performance in various domains, such as binary classification and regression problems, indicating their potential for deployment in energy forecasting, fault detection, predictive management, power system stability assessment, and data generation for power grid scenarios \cite{hong2023robust, Satpathy, khan, Chen2024, ajagekar2021, ajagekar2019, safari2024neuroquman, yu2023prediction, zhou2022noise, tang2022quantum, hangun2024SmartGrid, hangun2024wind}. Despite these advancements, challenges remain in implementing QML for energy forecasting. The currently available NISQ hardware imposes limitations on the scalability and performance of QML models. Efforts to mitigate noise and optimize quantum circuits are crucial to achieving reliable results in real-world applications. Consequently, further research into hyperparameter tuning and the robustness of QNN configurations across diverse scenarios is necessary to fully realize their potential.

\section{Methodology}
\subsection{Dataset}
Empirical analysis was conducted using a temporal dataset from \cite{DTU_Risoe_WindDatabase}, comprising 4,464 observations recorded at 10-minute intervals. The measurement parameters include wind velocity ($m/s$), directional component ($\theta$), atmospheric pressure ($hPa$), ambient temperature ($\degree C$), and power generation output ($kW$) of a wind turbine. In this study, we aimed to predict the power generation output using the rest of features. The target feature power generation output has about 2031 kW range. This wide range of values in this dataset makes it particularly suitable for evaluating QNN performance. For the experiments, the dataset was divided into 80\% training data and 20\% test data.

\subsection{Proposed Experiment Pipeline}
\begin{figure}
    \centering
    \includegraphics[width=0.4\textwidth]{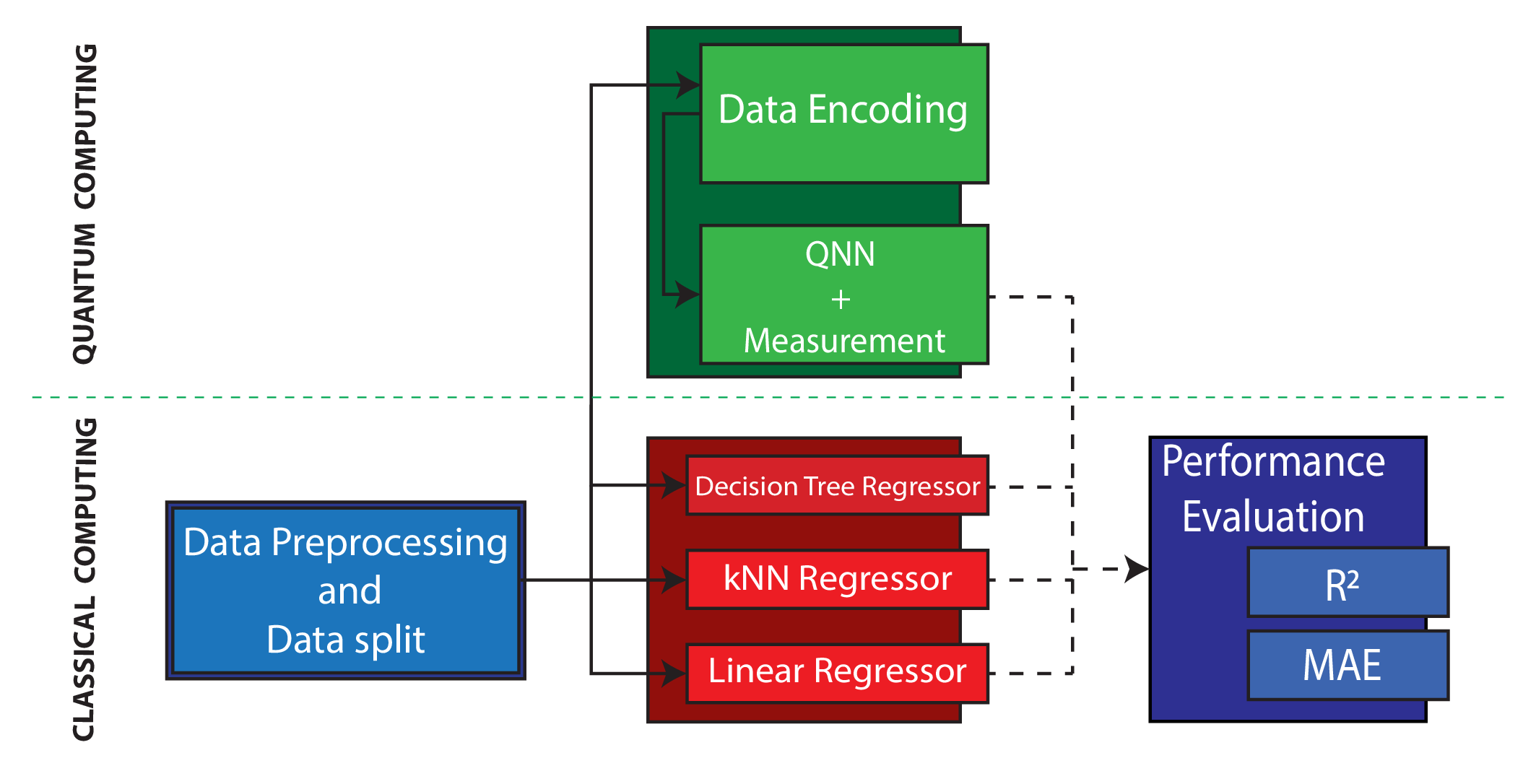}
    \caption{Pipeline of the experimental approach used in this study.}
    \label{fig:qnn_pipeline}
\end{figure}

We followed a three-step pipeline depicted in Fig. \ref{fig:qnn_pipeline}: data preprocessing and splitting, training classical and QNN models, and evaluating their performance. For the QNN models, this process involved preprocessing the data, encoding the data into a quantum representation, training the QNNs, and evaluating the trained QNNs using \(R^2\) and \(MAE\).

\subsection{Quantum Neural Networks (QNN)}
QNNs are special quantum circuits consisting of variational quantum circuits (VQCs) with tunable parameters. A QNN consists of three quantum circuits, which are a feature map, ansatz, and measurement. 

\subsubsection{Feature Map}
To encode classical data into the quantum domain (as shown in Fig. \ref{fig:qnn_pipeline}), a quantum circuit known as the \textit{feature map} is used. Common approaches for feature maps to encode classical data include \textit{angle encoding}, \textit{amplitude encoding}, or \textit{basis encoding}. These approaches essentially use rotation angles or amplitudes of basis states to represent classical data in the quantum domain.

\subsubsection{Ansatz}
In the context of quantum computing, an ansatz is a trial wave function or state that serves as an initial approximation to the solution of a given quantum problem, such as determining the ground state energy of a many-body system. Formally, it is typically represented as a parameterized quantum circuit, where adjustable parameters define the quantum state being approximated. By systematically varying these parameters and employing either classical or quantum optimization techniques, one aims to refine the ansatz until it closely approximates the target quantum state. 

\subsubsection{Measurement}
To obtain the prediction results from the QNN, and to convert the outputs of the quantum circuit into classical information, a measurement process is required. In this study, the expected value of all qubits is measured in the Pauli Z basis, and the results are returned in array format.

For the QML experiments, we evaluated twelve QNN configurations whose feature map and ansatz settings were given in Table \ref{tab:qnn_config}, and for the classical ML experiments, we decided to use kNN regressor, linear regressor, and decision tree regressor, whose prediction capabilities are proven in one of our previous works \cite{eyecioglu2019}.

\begin{table}
\centering
\caption{QNN CONFIGURATIONS USED IN EXPERIMENTS}
\label{tab:qnn_config}
\begin{tabular}{lll}
\toprule
QNN Configuration & Feature Map & Ansatz \\ 
\midrule
QNN-1  & Z Feature Map & Linear \\
QNN-2  & Z Feature Map & Full \\
QNN-3  & Z Feature Map & Circular \\
QNN-4  & Z Feature Map & SCA \\
QNN-5  & Z Feature Map & Reverse Linear \\
QNN-6  & Z Feature Map & Pairwise \\
QNN-7  & ZZ Feature Map & Linear \\
QNN-8  & ZZ Feature Map & Full \\
QNN-9 & ZZ Feature Map & Circular \\
QNN-10 & ZZ Feature Map & SCA \\
QNN-11 & ZZ Feature Map & Reverse Linear \\
QNN-12 & ZZ Feature Map & Pairwise \\ 
\bottomrule
\end{tabular}
\end{table}

\subsection{Evaluation}
\begin{figure}[h!]
    \centering
    \begin{subfigure}[]{0.4\textwidth}  
        \centering
        \includegraphics[width=\linewidth]{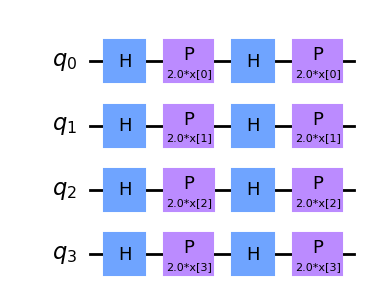}
        \caption{Single-qubit Z Feature Map for Data Encoding}  
        \label{fig:zfeaturemap_circ}
    \end{subfigure}
    \hfill
    \begin{subfigure}[]{0.48\textwidth} 
        \centering
        \includegraphics[width=\linewidth]{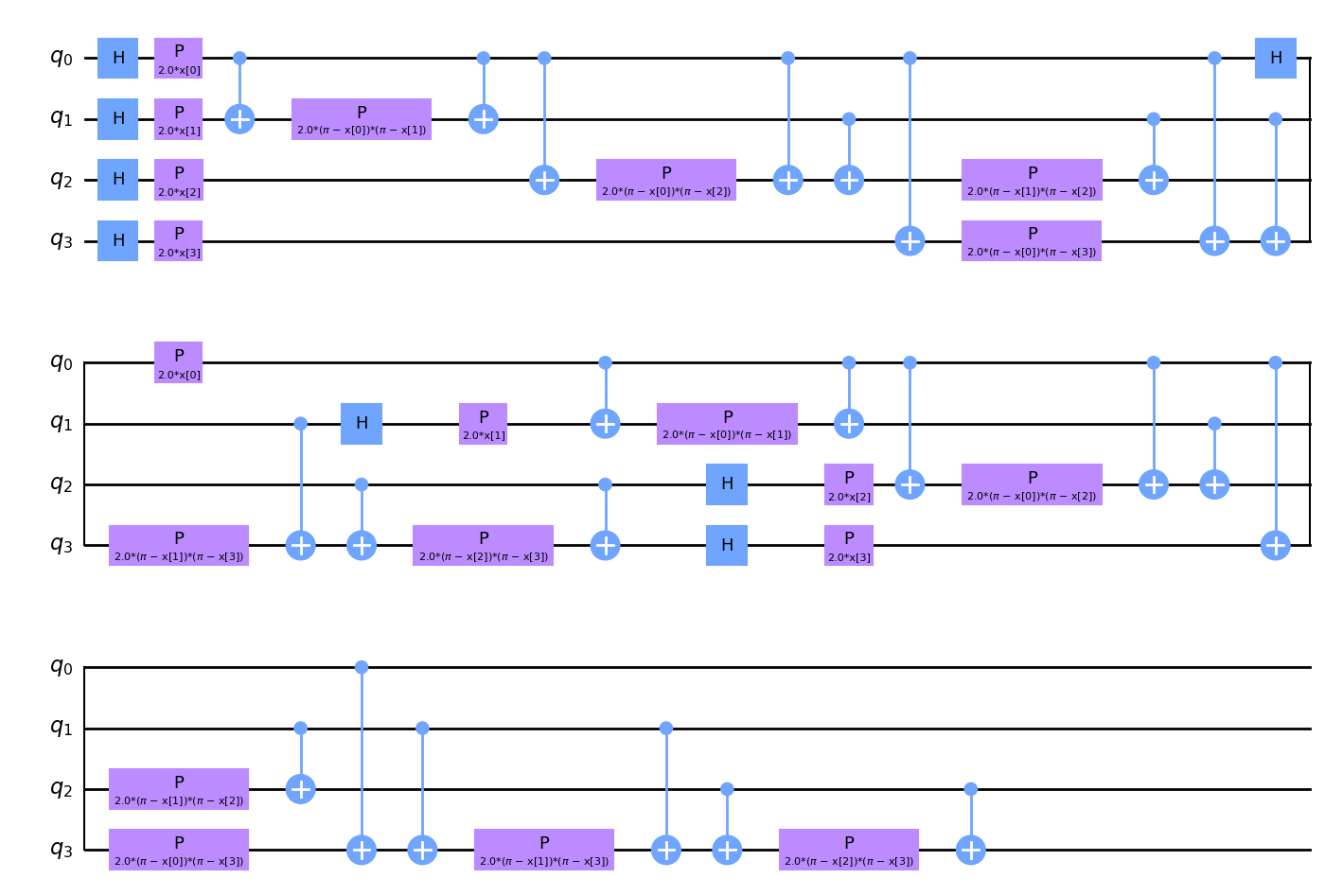}
        \caption{Two-qubit ZZ Feature Map for Data Encoding}  
        \label{fig:zzfeaturemap_circ}
    \end{subfigure}
    \caption{Quantum circuit implementations of Z and ZZ feature maps.}
    \label{fig:featuremaps}
\end{figure}

\begin{figure}[h!]
    \centering
    \begin{subfigure}[b]{0.32\textwidth}  
        \centering
        \includegraphics[width=\linewidth]{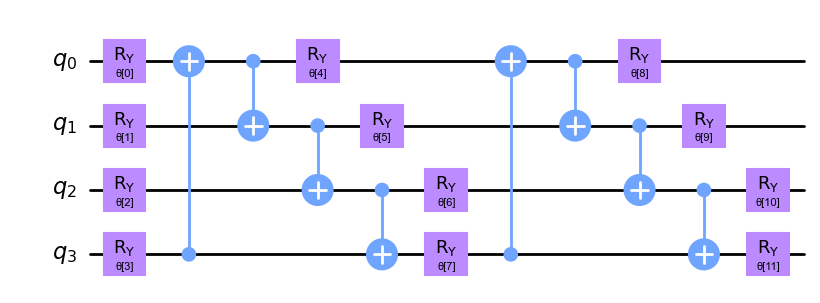}
        \caption{Circular Entanglement Architecture}  
        \label{fig:ansatz_circular}
    \end{subfigure}
    \hfill
    \begin{subfigure}[b]{0.32\textwidth}
        \centering
        \includegraphics[width=\linewidth]{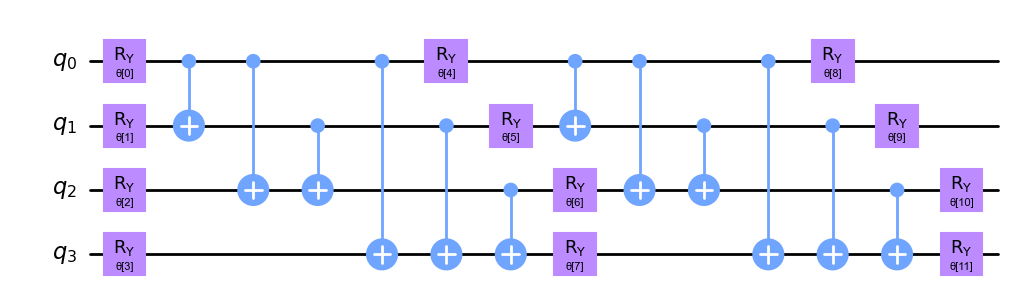}
        \caption{Full Entanglement Architecture}
        \label{fig:ansatz_full}
    \end{subfigure}
    \hfill
    \begin{subfigure}[b]{0.32\textwidth}
        \centering
        \includegraphics[width=\linewidth]{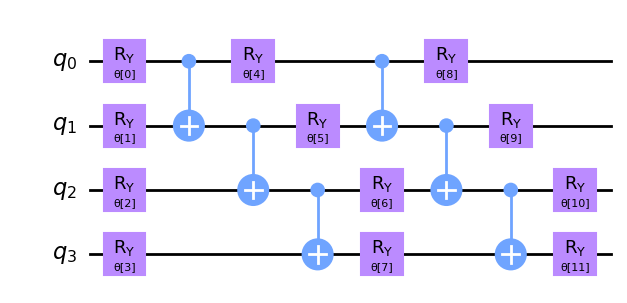}
        \caption{Linear Entanglement Architecture}
        \label{fig:ansatz_linear}
    \end{subfigure}

    \vspace{0.5cm}
    
    \begin{subfigure}[b]{0.32\textwidth}
        \centering
        \includegraphics[width=\linewidth]{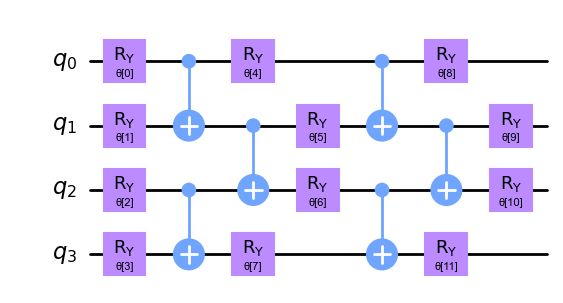}
        \caption{Pairwise Entanglement Architecture}
        \label{fig:ansatz_pairwise}
    \end{subfigure}
    \hfill
    \begin{subfigure}[b]{0.32\textwidth}
        \centering
        \includegraphics[width=\linewidth]{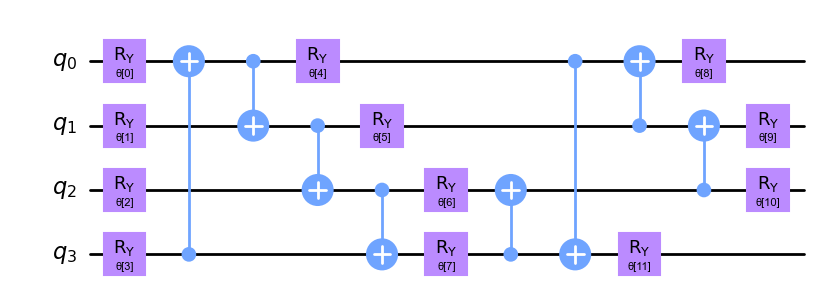}
        \caption{SCA Entanglement Architecture}
        \label{fig:ansatz_sca}
    \end{subfigure}
    \hfill
    \begin{subfigure}[b]{0.32\textwidth}
        \centering
        \includegraphics[width=\linewidth]{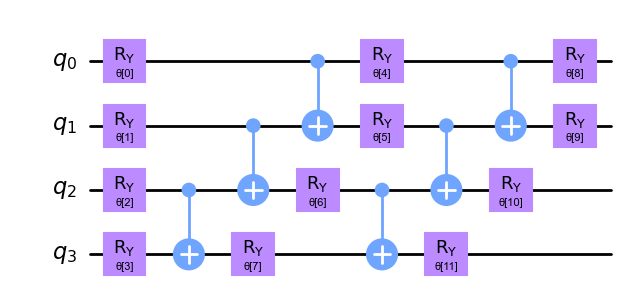}
        \caption{Reverse Linear Architecture}
        \label{fig:ansatz_reverse_linear}
    \end{subfigure}
    
    \caption{Variational circuit architectures with different entanglement strategies for state preparation.}
    \label{fig:ansatz_patterns}
\end{figure}

In this study, we aimed to provide a comparative approach to the effect of the QNN structure on prediction performance. We evaluated twelve QNN configurations, combining two feature maps (Z and ZZ) with six ansatz circuits. Our initial feature map selection was the Z feature map, whose quantum circuit representation (Fig. \ref{fig:zfeaturemap_circ}) comprises Hadamard ($H$) gates and Phase ($P$) gates. For the second set of experiments, we used the ZZ feature map, whose quantum circuit representation is given in Fig. \ref{fig:zzfeaturemap_circ} which consists of H gates, P gates, and CNOT($CX$) gates. Finally, for the ansatz, we constructed six distinct circuits based on different qubit entanglement strategies. We used linear entanglement, circular entanglement, full entanglement, reverse linear entanglement, pairwise entanglement, and shifted-circular-alternating (SCA) entanglement. The description of all ansatz circuits is given in Fig. \ref{fig:ansatz_patterns}. Depending on the entanglement strategy, the ansatz circuits used in experiments contain numerous gates, which can introduce errors into the calculation and/or can lead the barren plateau problems. While this study acknowledges this potential source of limitations, a detailed experimental investigation of its effects is reserved for future work. To simulate given QNN configurations, we used Qiskit \cite{qiskit2024} and Qiskit Machine Learning \cite{qiskitml2025} with Python. As the classical optimizer part of the QNN, we used the Limited-memory Broyden-Fletcher-Goldfarb-Shanno Bound (L-BFGS-B) optimizer with 25 iterations, and $\epsilon$(error) of $1e-8$. The value of $\epsilon$ is given by Qiskit's default. For this study, each circuit was run once to assess its single-run performance. A more detailed test, consisting of several runs, has been reserved for an extended version of this study. Both classical and quantum approaches were run on an HP Z1 G2 Workstation equipped with 16 GB of RAM and a 3.5 GHz CPU. For the quantum approaches, simulations showed that the average training execution time was approximately 44 minutes for circuits based on the Z feature map, and approximately 88 minutes for those based on the ZZ feature map.

\section{Results}
\begin{figure*}
    \centering
    \includegraphics[width=0.7\linewidth]{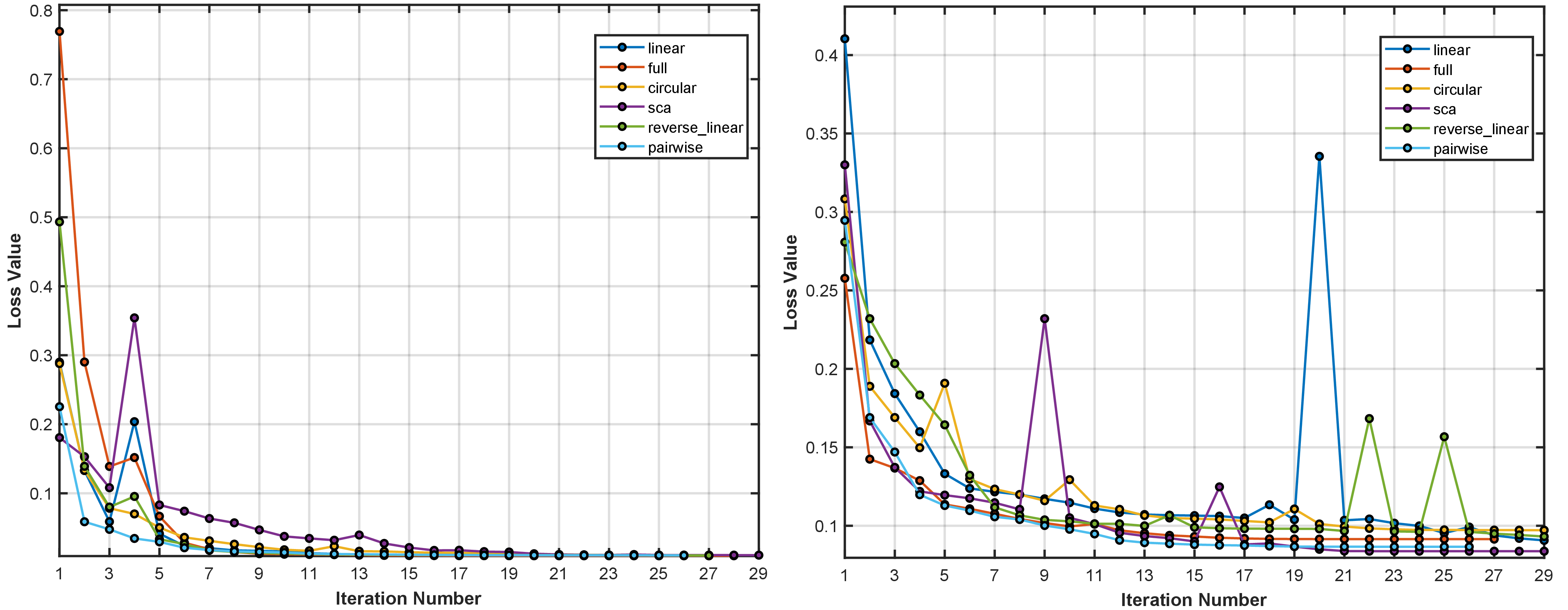}
    \caption{Learning process of QNNs for wind power prediction, comparing (left) Z feature map and (right) ZZ feature map.}
    \label{fig:all_object_func}
\end{figure*}

For comparison, we examined the learning behavior of these QNN configurations by observing the change in the objective function for each QNN circuit. Fig. \ref{fig:all_object_func} presents the learning behavior of QNN configurations based on the Z feature map and those based on the ZZ feature map. According to our results, QNN configurations employing the Z feature map minimize the objective function more effectively and exhibit fewer oscillations during convergence. Conversely, QNN configurations based on the ZZ feature map are less effective at minimizing the objective function, and oscillations are more prevalent. Although the maximum number of iterations was set to 25, Fig. \ref{fig:all_object_func} demonstrates that Z feature map-based QNN configurations converge to a sufficiently low loss value in fewer iterations. This indicates that identifying an optimal pairing of total iteration count and target error is important for achieving comparable learning performance in fewer iterations.

\begin{table}
\centering
\caption{PERFORMANCE COMPARISON BETWEEN QUANTUM NEURAL NETWORKS AND CLASSICAL METHODS}
\label{tab:perf_metrc_comp}
\begin{tabular}{llcc}
\hline
\textbf{Method} & \textbf{Approach} & \textbf{R\textsuperscript{2}} & \textbf{MAE} \\ \hline
& QNN-1 & 0.92 & 136.50 \\ \cline{2-4} 
& QNN-2 & \textbf{0.93} & 123.81 \\ \cline{2-4} 
& QNN-3 & \textbf{0.93} & 119.71 \\ \cline{2-4} 
& QNN-4 & 0.92 & 134.59 \\ \cline{2-4} 
& QNN-5 & \textbf{0.93} & 119.05 \\ \cline{2-4} 
\textbf{Quantum} & QNN-6 & 0.92 & 134.45 \\ \cline{2-4} 
& QNN-7 & 0.35 & 446.02 \\ \cline{2-4} 
& QNN-8 & 0.34 & 462.55 \\ \cline{2-4} 
& QNN-9 & 0.29 & 478.16 \\ \cline{2-4} 
& QNN-10 & 0.33 & 443.65 \\ \cline{2-4} 
& QNN-11 & 0.34 & 462.29 \\ \cline{2-4} 
& QNN-12 & 0.34 & 440.12 \\ \hline
& Decision Tree & 0.91 & 66.38 \\ \cline{2-4} 
\textbf{Classical} & k-Nearest Neighbors & 0.92 & 103.30 \\ \cline{2-4} 
& Linear Regression & 0.88 & 162.76 \\ \hline
\end{tabular}
\end{table}

Although the training curves demonstrate the learning process, we can better assess the practical utility of the models through their predictive performance. In Fig. \ref{fig:act_vs_pred_fig}, the actual and predicted values for each QNN circuit are detailed. According to the results, all Z feature map-based QNN configurations yield $R^2$ values around 0.92-0.93, as shown in Fig. \ref{fig:sca_pred_zf}. On the other hand, none of the ZZ feature map-based QNN configurations achieved an acceptable performance.

As shown in Table \ref{tab:perf_metrc_comp}, the key performance metrics indicate the overall success of the Z feature map-based QNN configurations. We evaluated the models using two key metrics: coefficient of determination (R\textsuperscript{2}), mean absolute error (MAE). Performance comparison across different QNN configurations reveals several significant patterns. The Z feature map consistently shows superior performance compared to the ZZ feature map across all ansatz types, with R\textsuperscript{2} values ranging from 0.92 to 0.93 for Z implementations versus 0.29 to 0.35 for ZZ variants. The reverse linear ansatz combined with the Z feature map emerges as the optimal configuration, achieving the highest R\textsuperscript{2} (0.93) and lowest error metric ( MAE=119.71). Although Z feature map implementations show relatively consistent performance across different types of ansatz, the ZZ feature map exhibits greater variability, with the linear ansatz achieving the best performance (R\textsuperscript{2}= 0.35) between ZZ implementations and the circular ansatz showing the poorest performance (R\textsuperscript{2}= 0.29).

Based on comprehensive experimental results, Z feature map-based QNN configurations successfully solve prediction problems. Nevertheless, the current limitations of NISQ-era quantum computing continue to prompt skepticism regarding its practical applications. To overcome this skepticism, quantum computing researchers must demonstrate that quantum computing-based solutions can compete effectively with classical computing approaches. We also compared our results with selected classical approaches that showed success in a previous work that uses the same dataset \cite{eyecioglu2019}. If we compare the prediction results of the classical methods given in Figs. \ref{fig:cls_dtr_pred}, \ref{fig:cls_knn_pred}, \ref{fig:cls_linear_pred}, it is clear that the QNN configurations based on the Z feature map outperform the decision tree regressor and linear regressor and also compete with the k nearest neighbor regressor in terms of $R^2$ score. However, the decision tree regressor achieved lower statistical errors (MAE) compared to all QNN configurations. This indicates that while QNNs are more effective at capturing overall trends and reducing prediction variance, the decision tree regressor maintains better absolute accuracy. Overall, our experimental results suggest that quantum computing-based machine learning can be competitive with classical approaches.

\section{Conclusion}
In this experimental study, our aim was to investigate the effects of the different QNN configurations' success rates on a prediction problem and to provide a possible road map for other researchers in the field of QML for their future studies. Our experiments offer detailed insights into how various QNN configurations influence prediction performance.

We evaluated the functional performance of twelve different QNN configurations. The combinations of QNN configurations that use the Z feature map clearly showed their superiority over QNN configurations that use the ZZ feature map. Finally, we compared successful Z feature map-based QNN configurations with classical approaches to evaluate their competitiveness with already deployed methods. 
Experiments demonstrated that QNN configurations could compete effectively with classical methods, even surpassing them in some cases. Currently, QML is still developing, lacking widely accepted implementations compared to more mature classical machine learning approaches. We believe that this experimental study will contribute to QML on its way to mature implementation and generalization. Even though this study compares QNN configurations and their performance on prediction tasks, an in-depth evaluation is needed to assess from a broader perspective. As future work, we will extend this study to consider time complexity, data size, and time evaluation to provide a deeper and more detailed experiment to draw a bigger picture that can help QML researchers choose QNN configurations tailored to their needs.

\bibliographystyle{IEEEtran}
\bibliography{main}

\clearpage

\onecolumn

\vspace*{1cm}
\begin{center}
\Large\textbf{APPENDIX}
\end{center}
\vspace{0.5cm}

\section*{Circuit Diagrams}
\label{appendix:circuits}

\newpage

\section*{Performance Graphs}
\label{appendix:performance_graphs}
\begin{samepage} 
\vspace{-0.25cm}

\begin{figure}[H] 
\centering
\begin{subfigure}[b]{0.32\textwidth}
\centering
\includegraphics[width=0.8\linewidth]{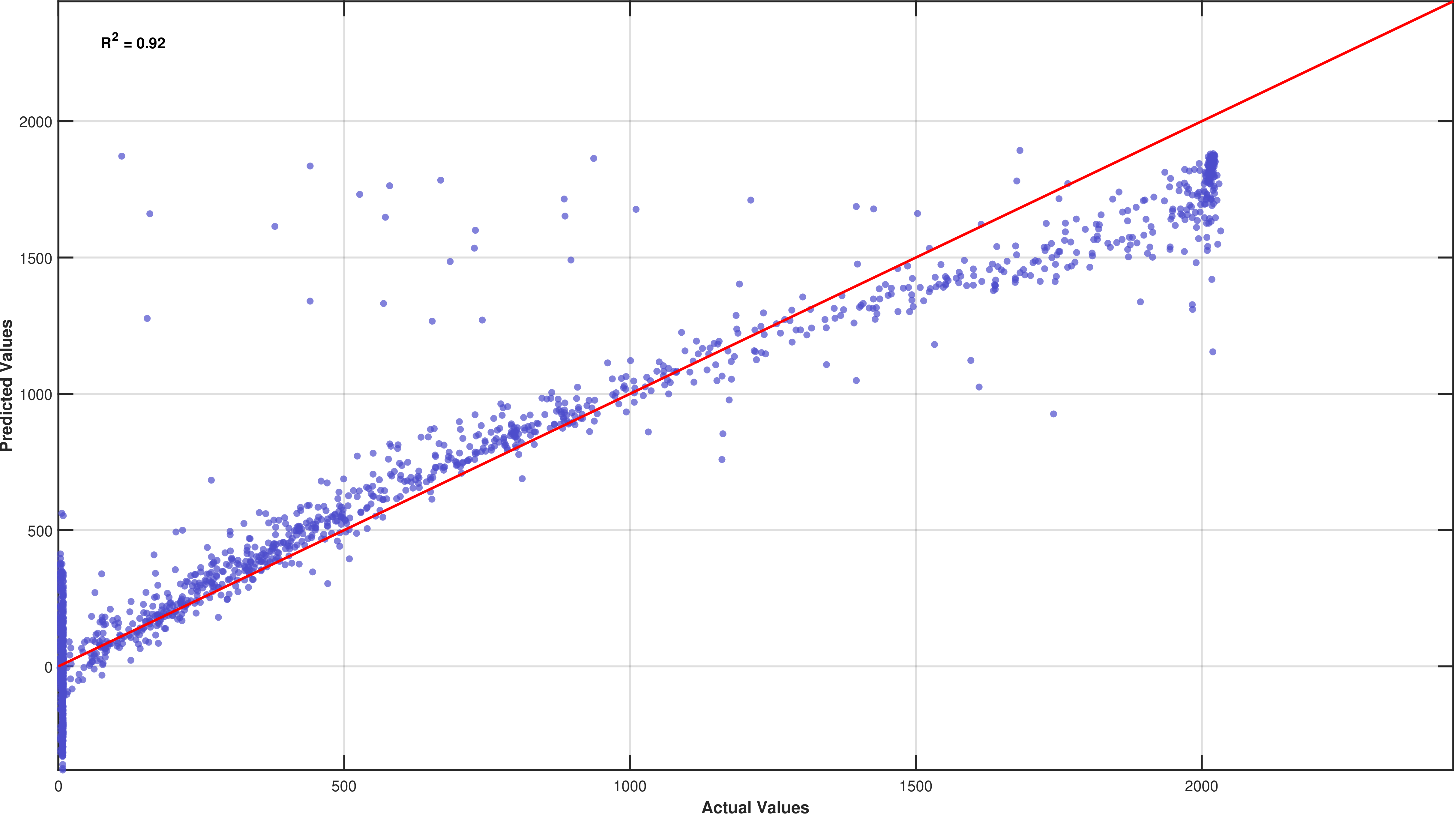}
\caption{QNN-1}
\label{fig:linear_pred_zf}
\end{subfigure}
\hfill
\begin{subfigure}[b]{0.32\textwidth}
\centering
\includegraphics[width=0.8\linewidth]{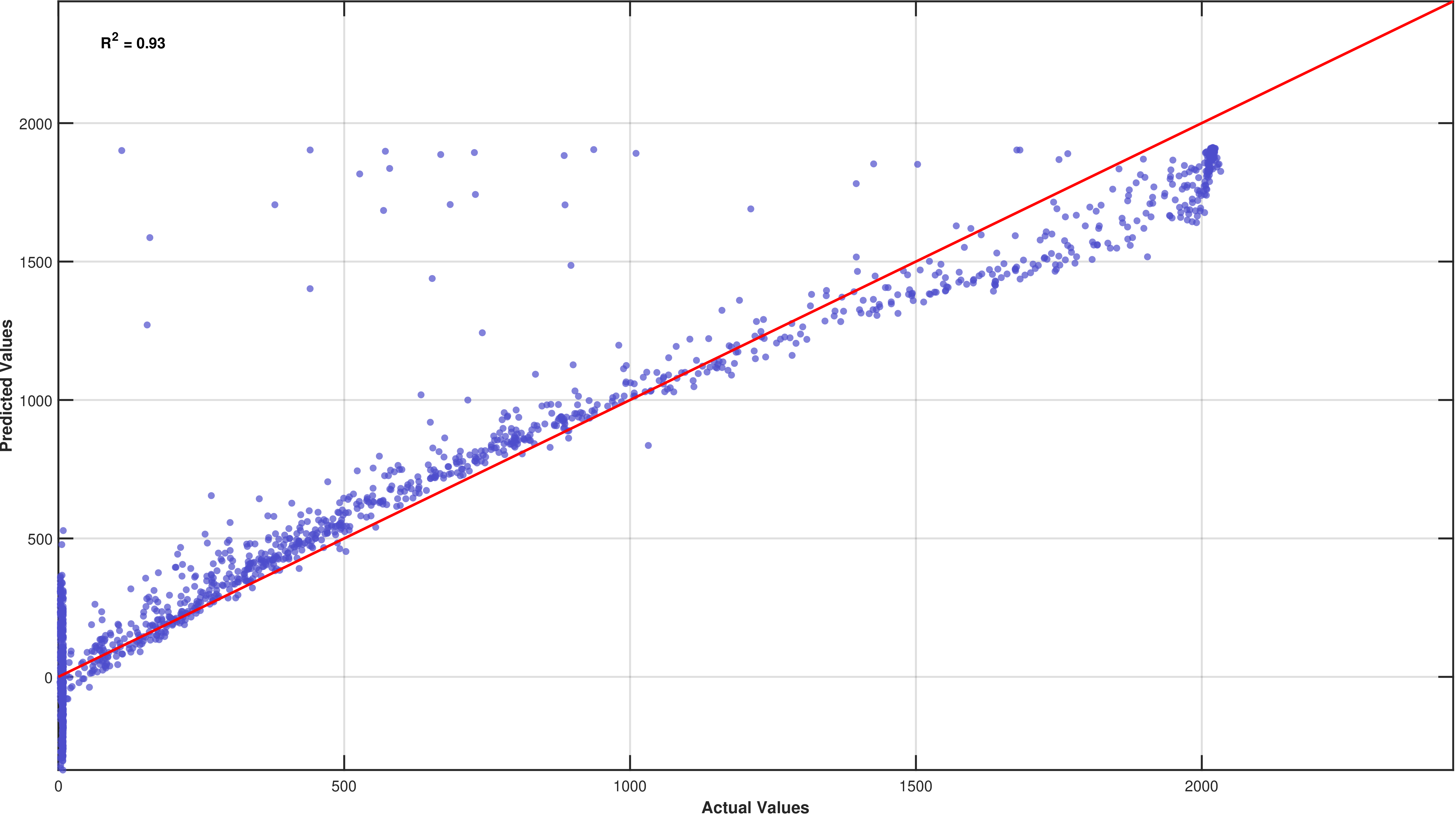}
\caption{QNN-2}
\label{fig:full_pred_zf}
\end{subfigure}
\hfill
\begin{subfigure}[b]{0.32\textwidth}
\centering
\includegraphics[width=0.8\linewidth]{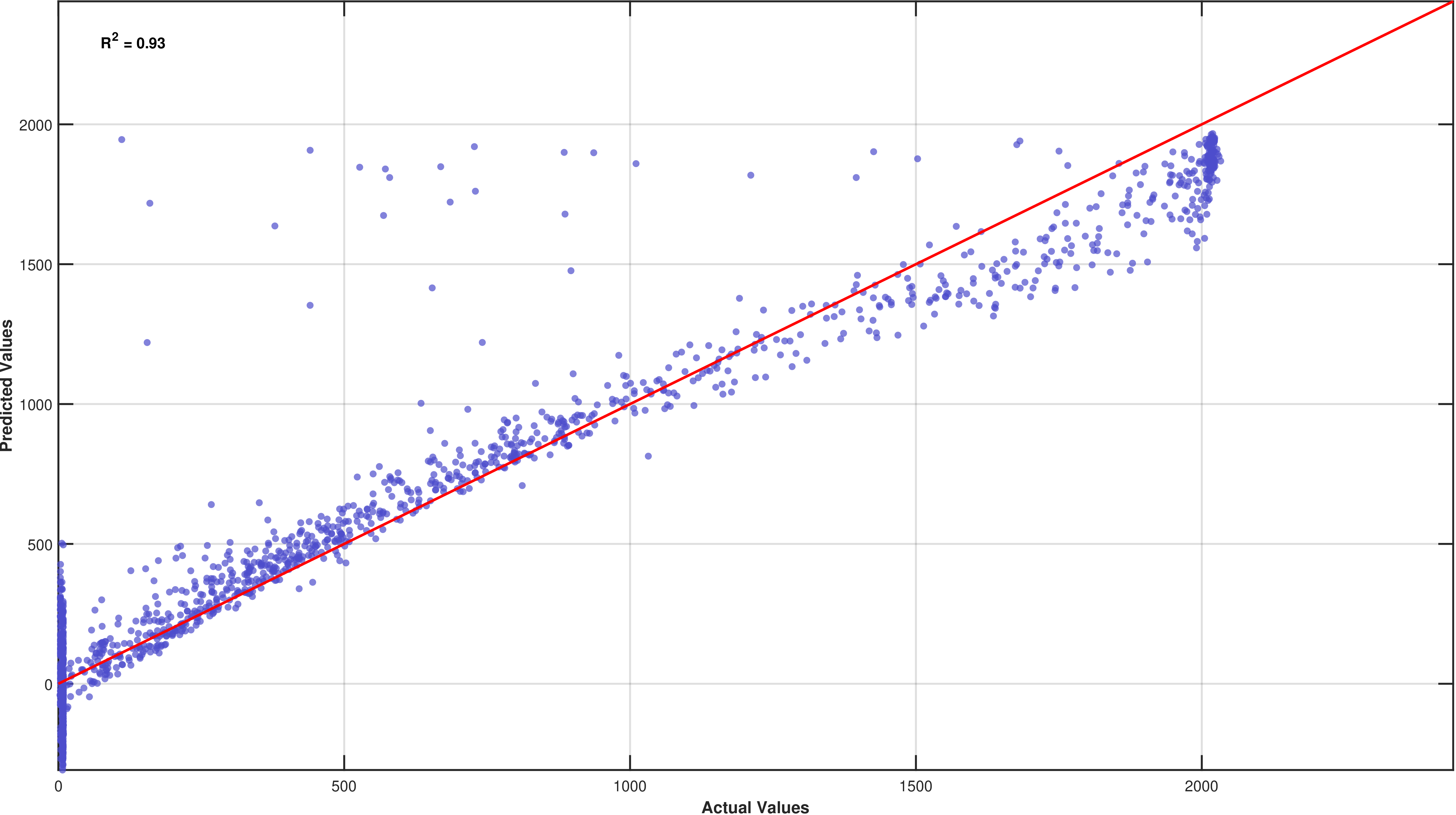}
\caption{QNN-3}
\label{fig:circ_pred_zf}
\end{subfigure}

\vspace{0.5cm}

\begin{subfigure}[b]{0.32\textwidth}
    \centering
\includegraphics[width=0.8\linewidth]{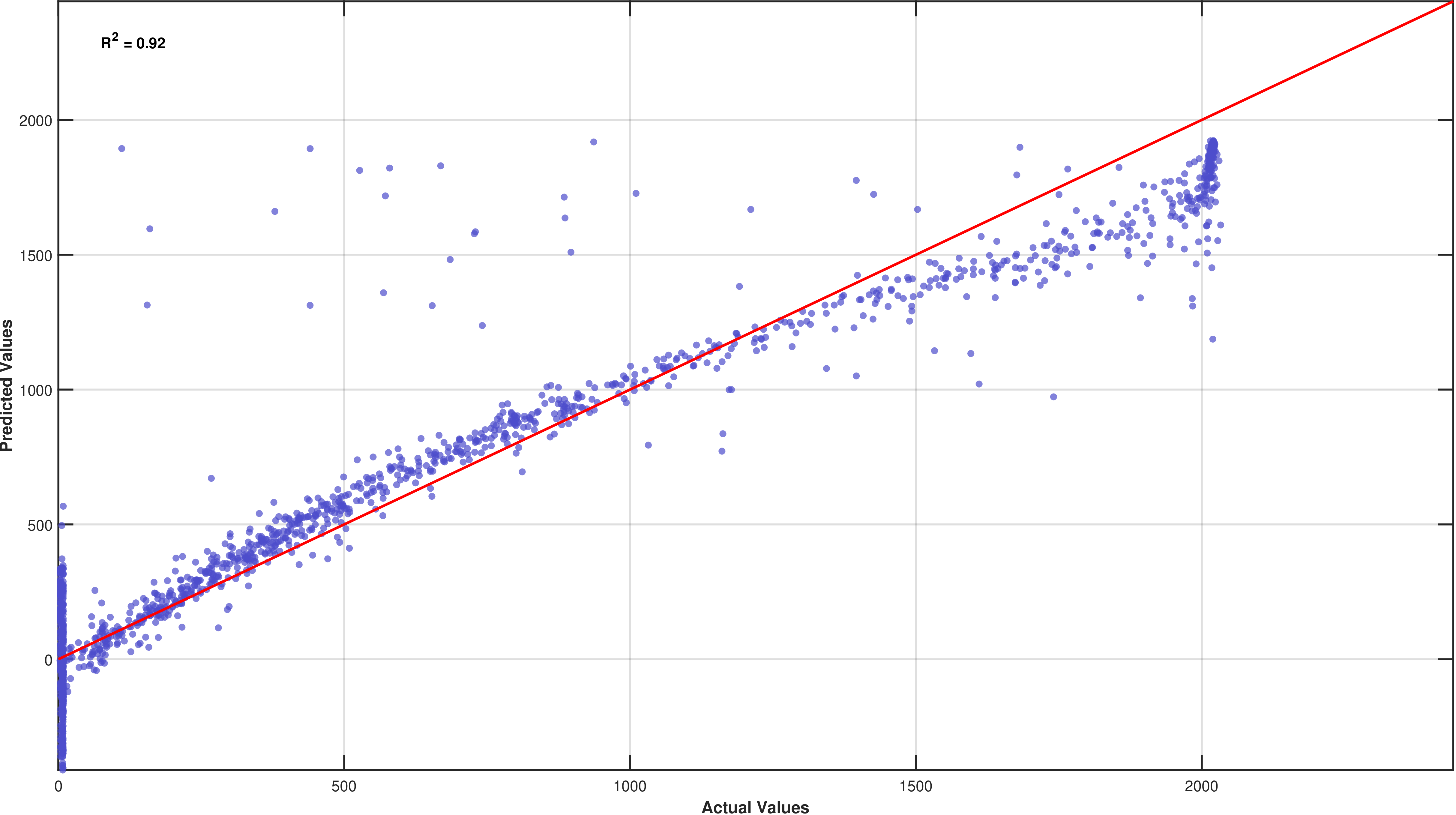}
    \caption{QNN-4}
    \label{fig:sca_pred_zf}
\end{subfigure}
\hfill
\begin{subfigure}[b]{0.32\textwidth}
    \centering
\includegraphics[width=0.8\linewidth]{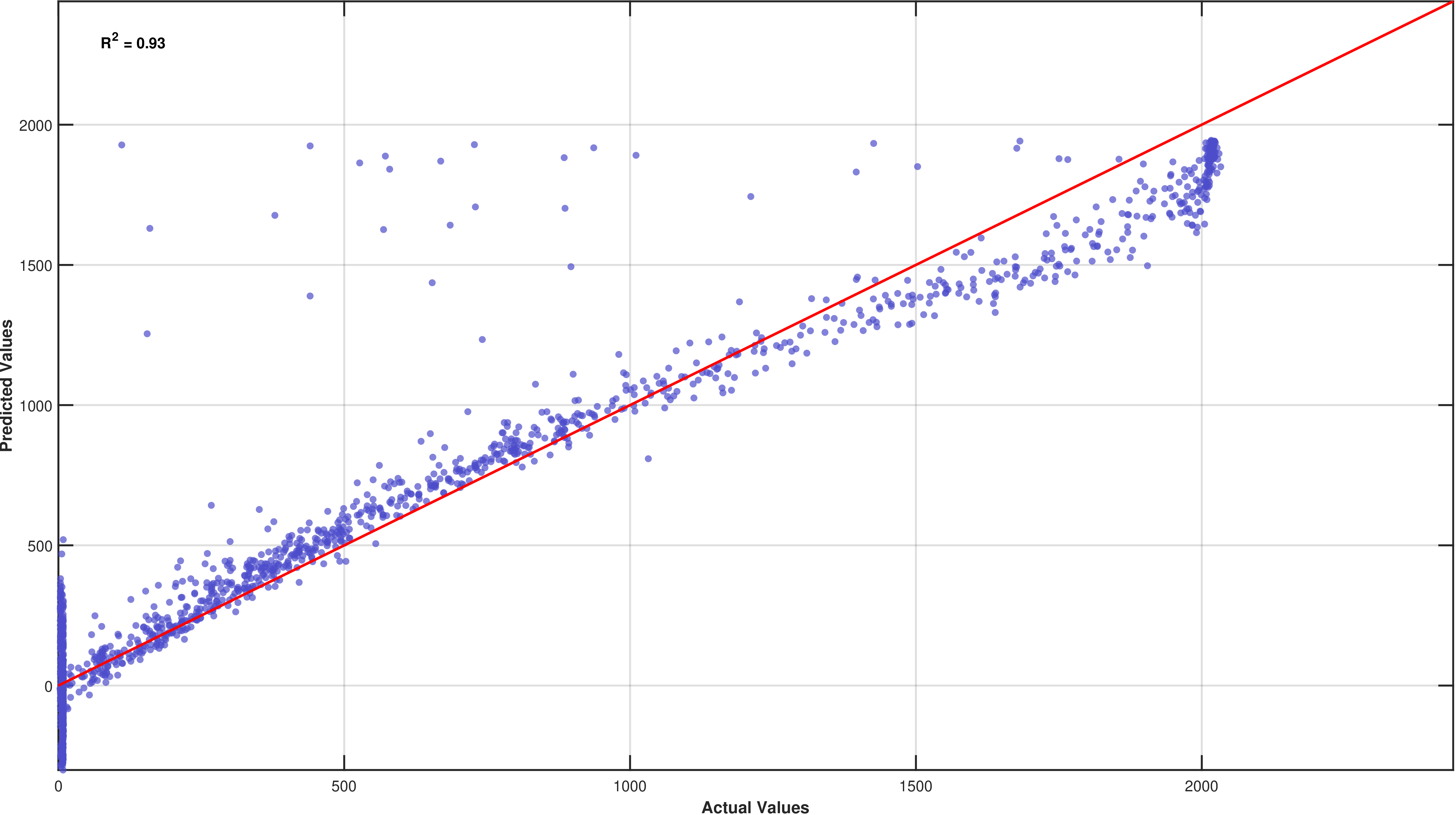}
    \caption{QNN-5}
    \label{fig:reverse_linear_pred_zf}
\end{subfigure}
\hfill
\begin{subfigure}[b]{0.32\textwidth}
    \centering
\includegraphics[width=0.8\linewidth]{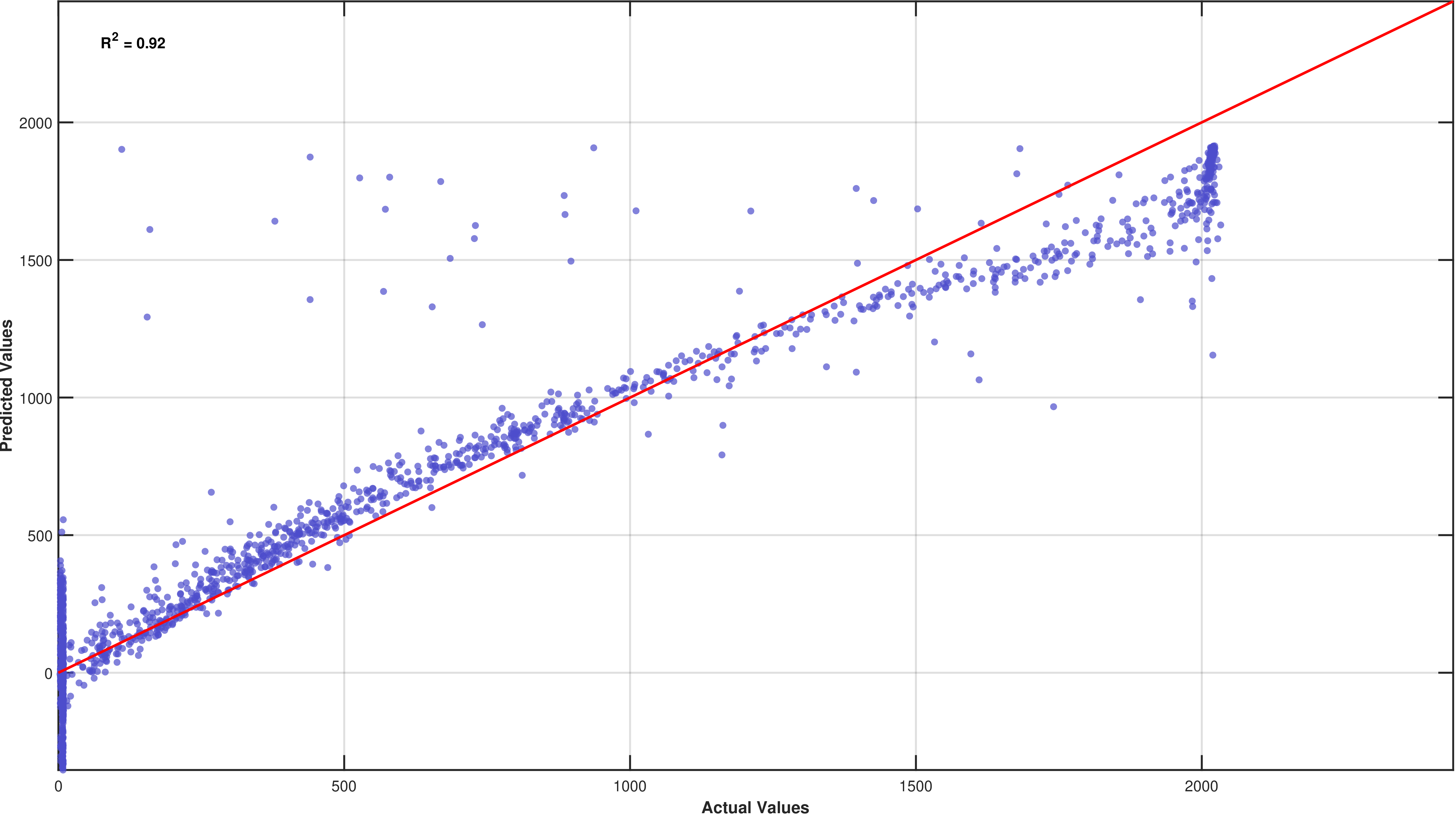}
    \caption{QNN-6}
    \label{fig:pairwise_pred_zf}
\end{subfigure}

\vspace{0.5cm}

\begin{subfigure}[b]{0.32\textwidth}
    \centering
\includegraphics[width=0.8\linewidth]{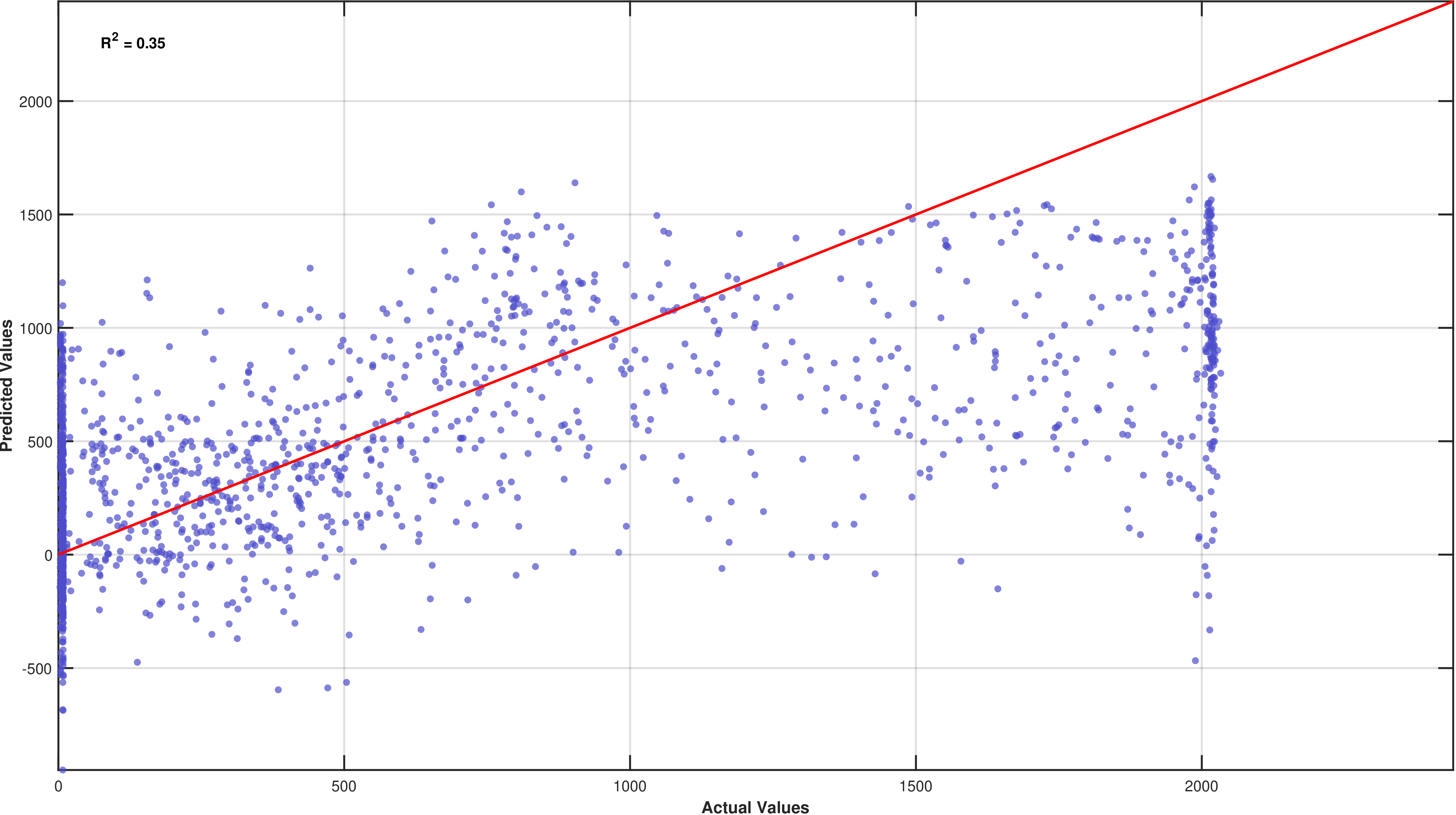}
    \caption{QNN-7}
    \label{fig:linear_pred_zzf}
\end{subfigure}
\hfill
\begin{subfigure}[b]{0.32\textwidth}
    \centering
\includegraphics[width=0.8\linewidth]{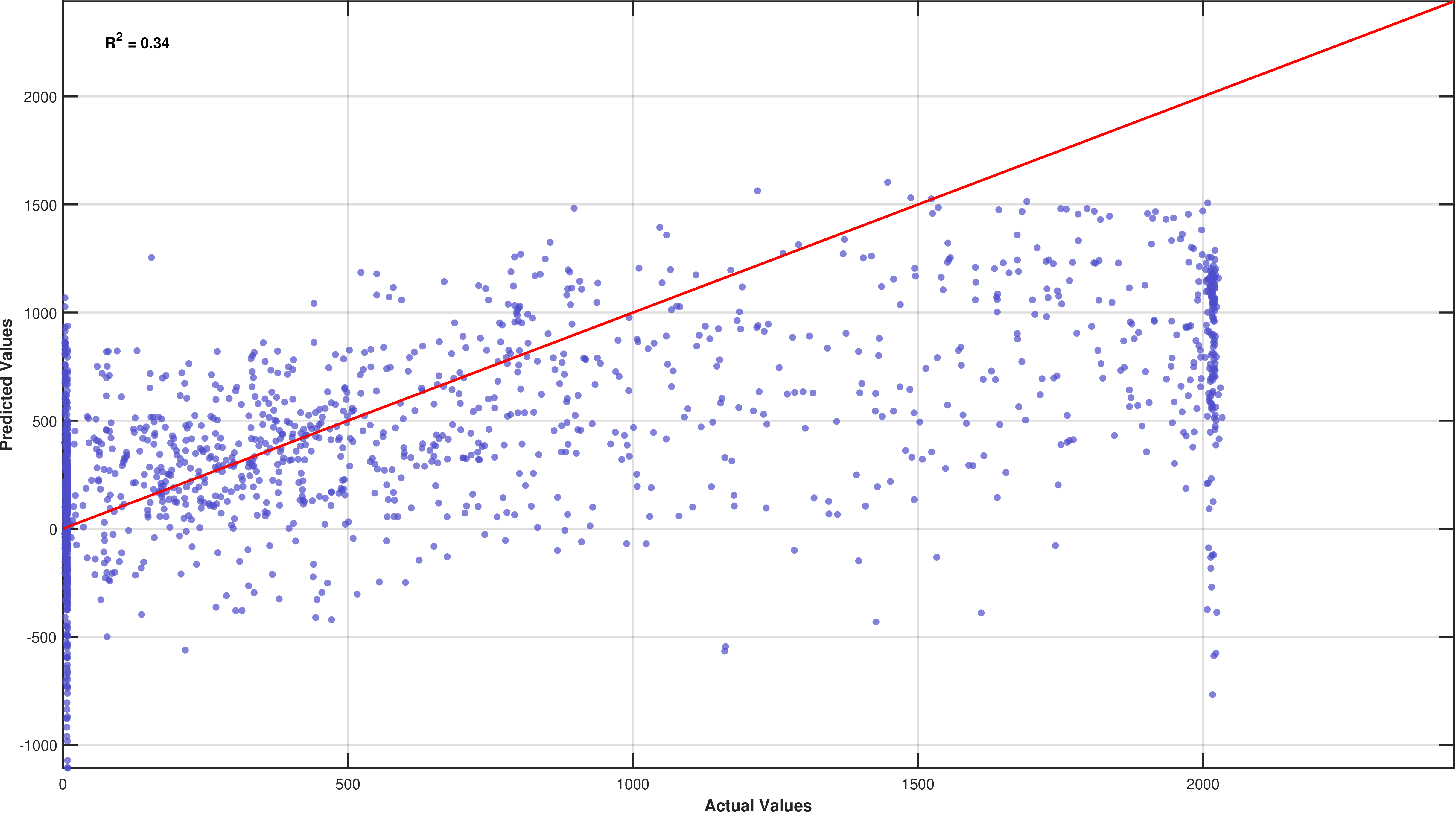}
    \caption{QNN-8}
    \label{fig:full_pred_zzf}
\end{subfigure}
\hfill
\begin{subfigure}[b]{0.32\textwidth}
    \centering
\includegraphics[width=0.8\linewidth]{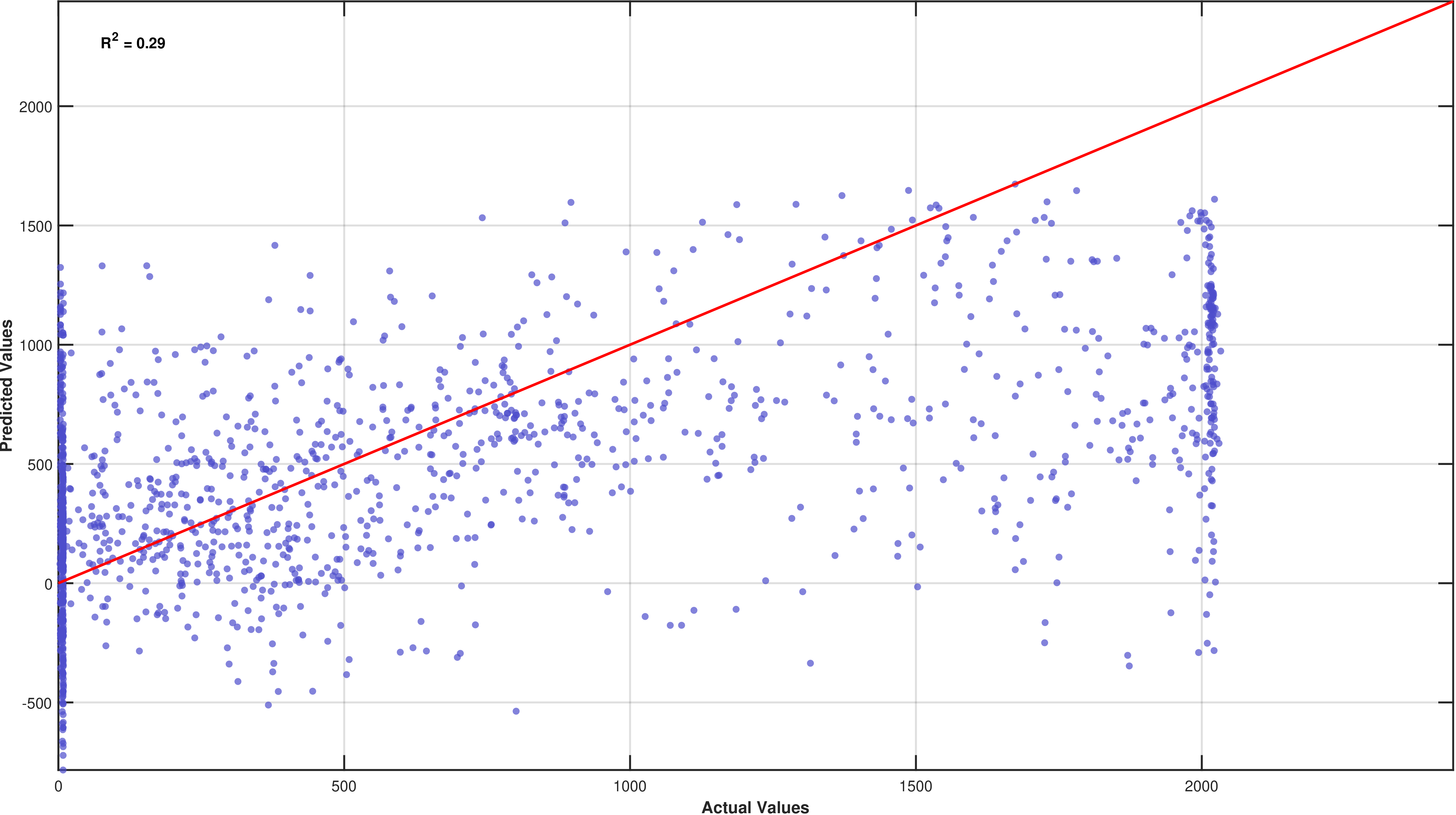}
    \caption{QNN-9}
    \label{fig:circ_pred_zzf}
\end{subfigure}

\vspace{0.5cm}

\begin{subfigure}[b]{0.32\textwidth}
    \centering
\includegraphics[width=0.8\linewidth]{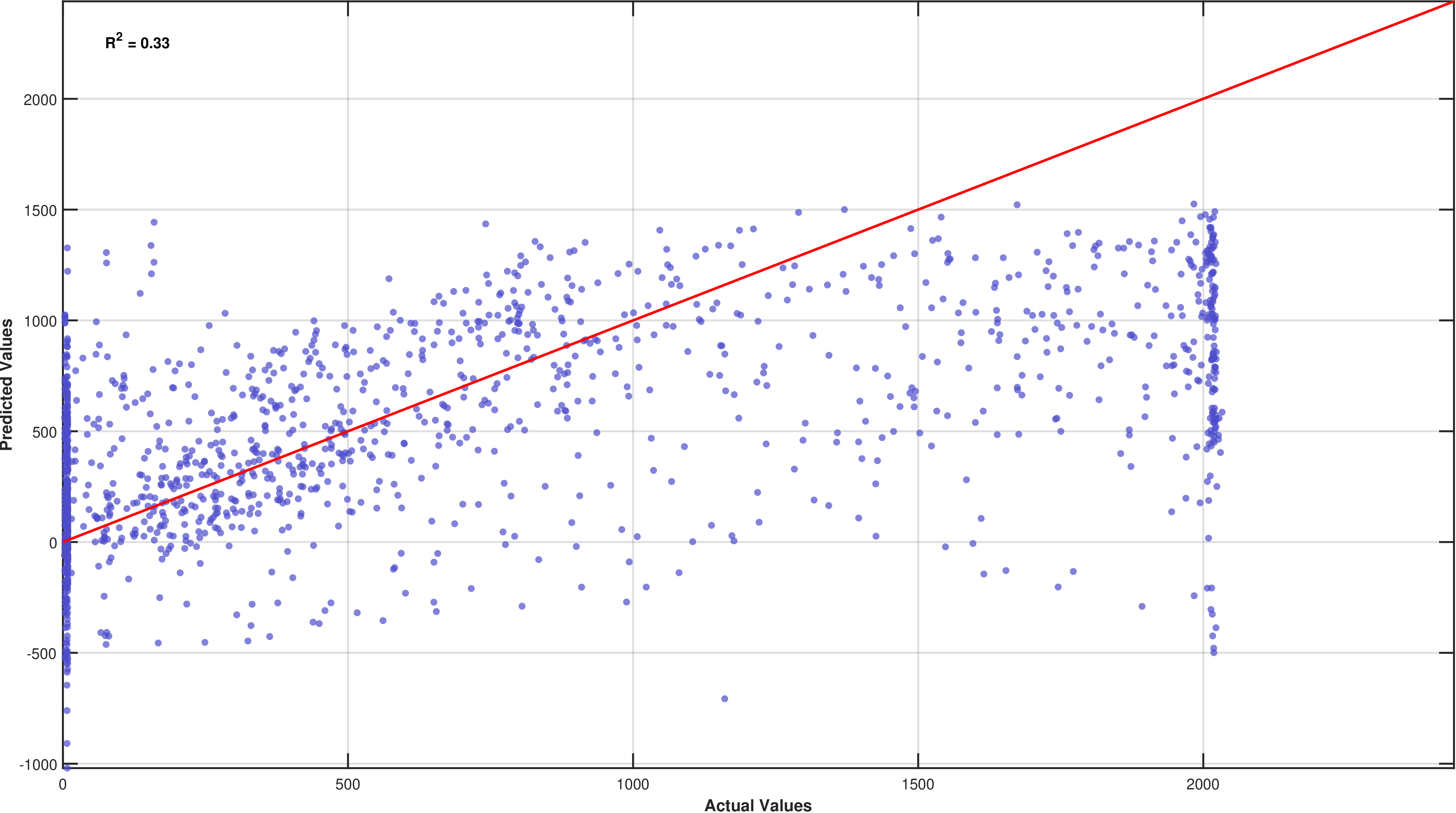}
    \caption{QNN-10}
    \label{fig:sca_pred_zzf}
\end{subfigure}
\hfill
\begin{subfigure}[b]{0.32\textwidth}
    \centering
\includegraphics[width=0.8\linewidth]{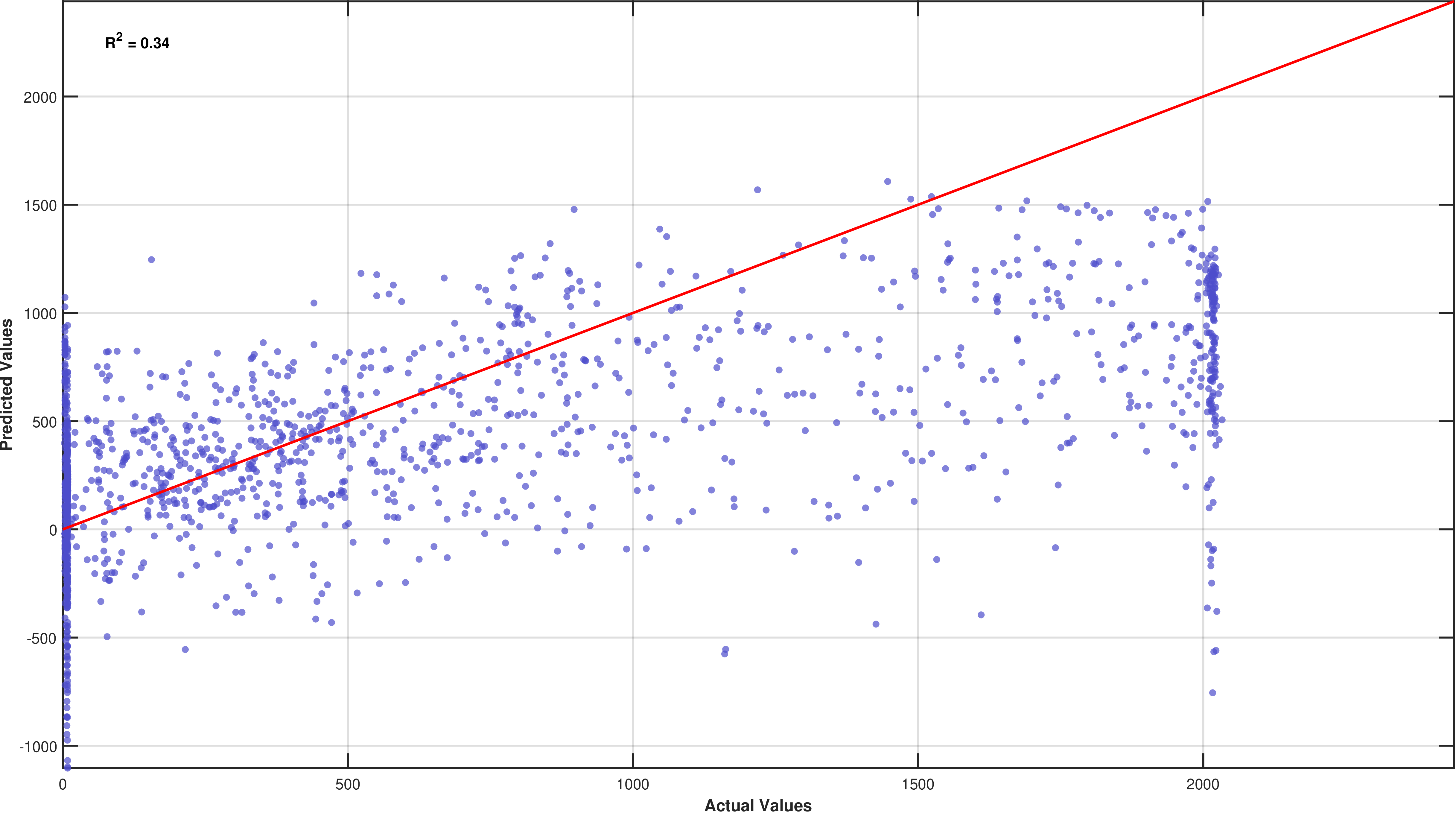}
    \caption{QNN-11}
    \label{fig:reverse_linear_pred_zzf}
\end{subfigure}
\hfill
\begin{subfigure}[b]{0.32\textwidth}
    \centering
\includegraphics[width=0.8\linewidth]{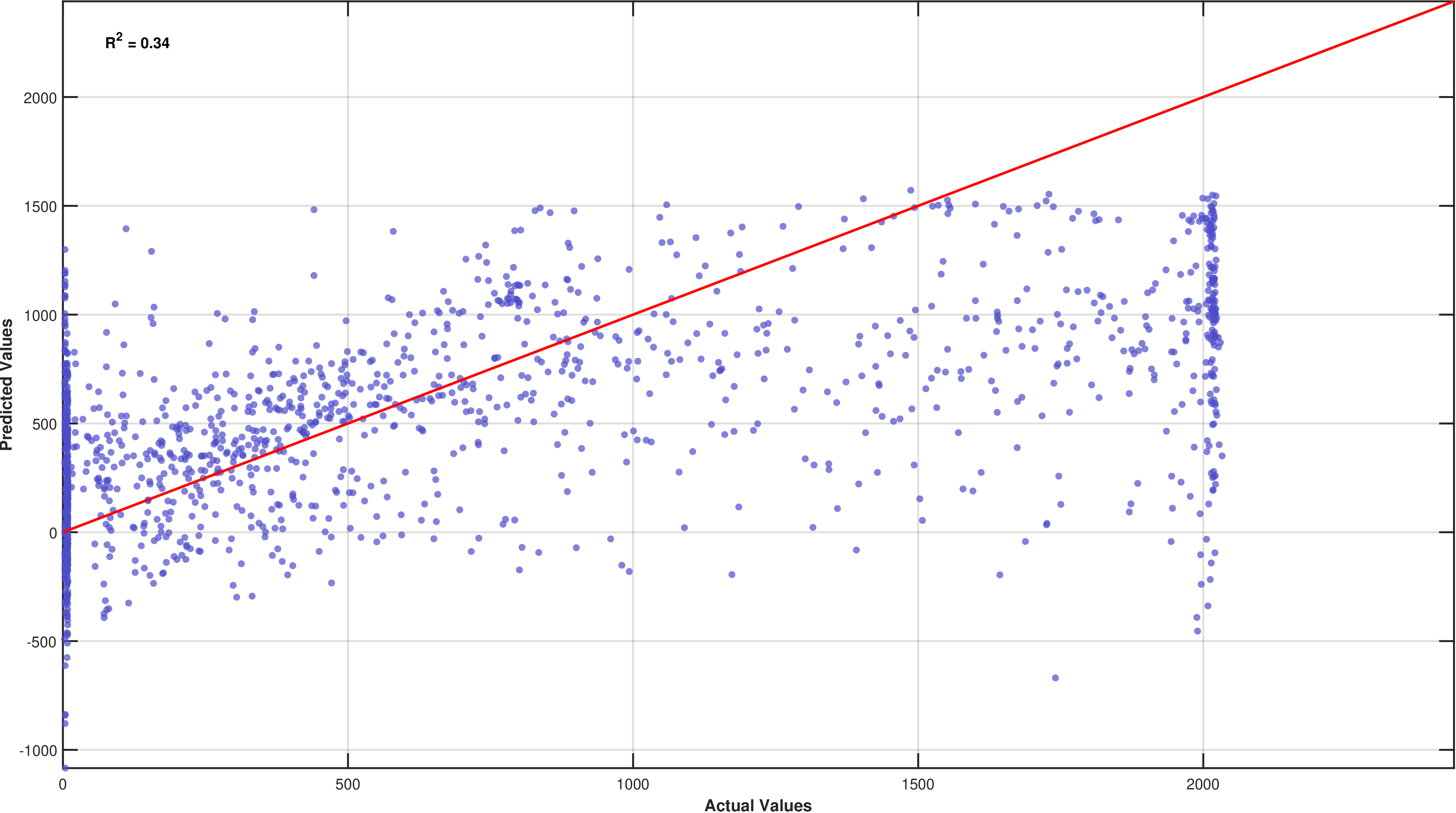}
    \caption{QNN-12}
    \label{fig:pairwise_pred_zzf}
\end{subfigure}

\vspace{0.5cm}

\begin{subfigure}[b]{0.32\textwidth}
    \centering
\includegraphics[width=0.8\linewidth]{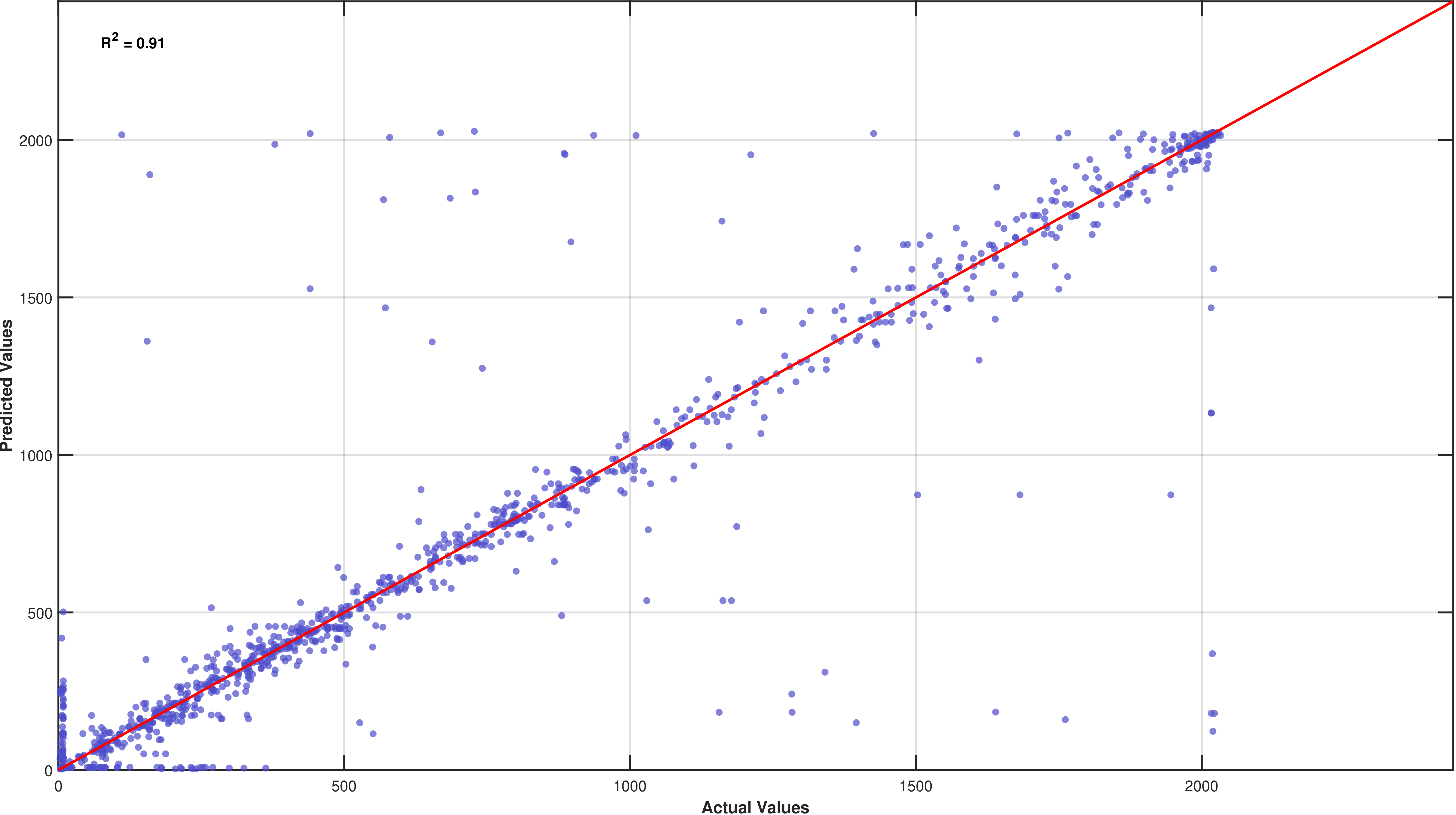}
    \caption{Decision Tree Regression}
    \label{fig:cls_dtr_pred}
\end{subfigure}
\hfill
\begin{subfigure}[b]{0.32\textwidth}
    \centering
\includegraphics[width=0.8\linewidth]{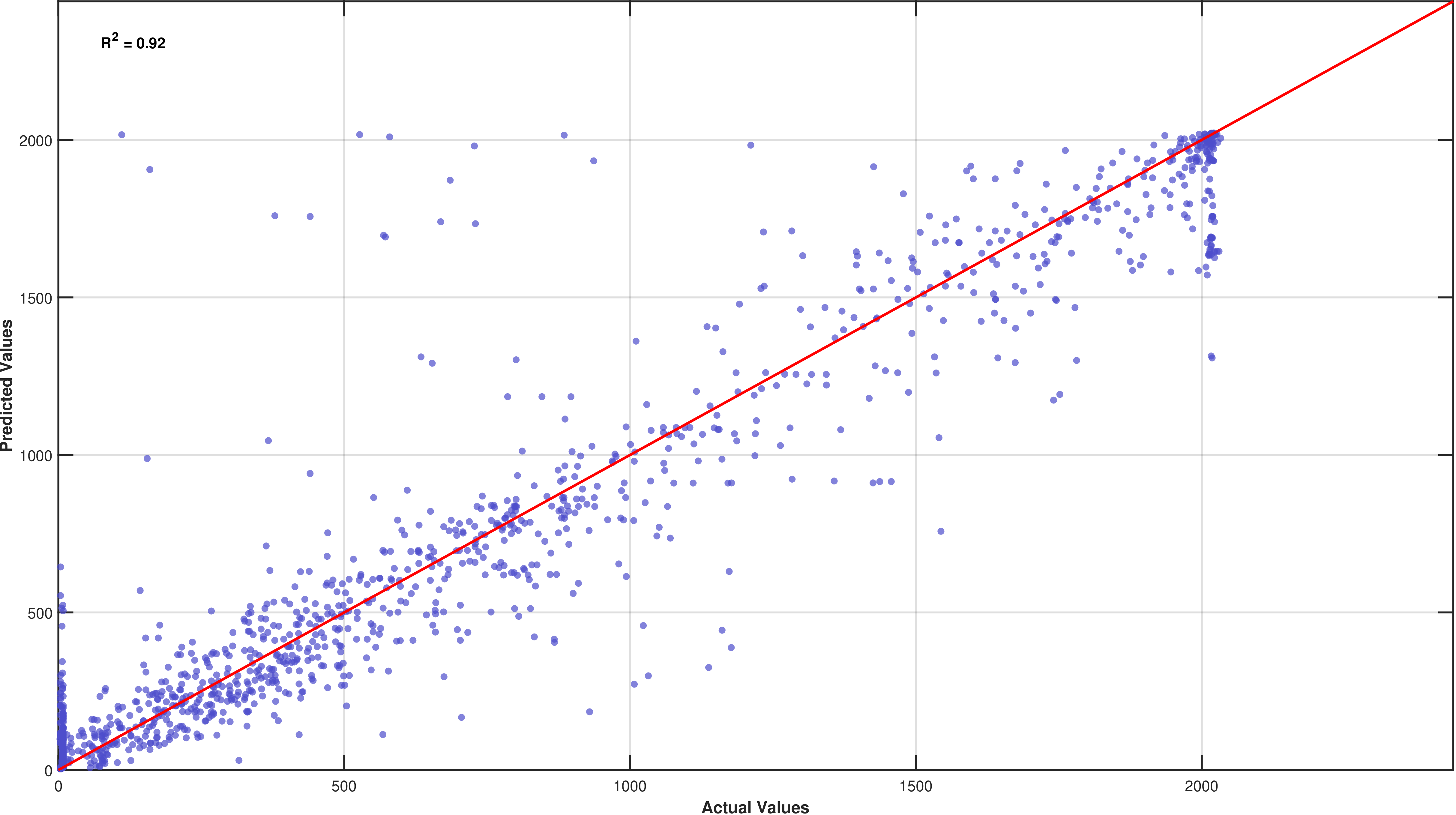}
    \caption{k-Nearest Neighbors}
    \label{fig:cls_knn_pred}
\end{subfigure}
\hfill
\begin{subfigure}[b]{0.32\textwidth}
    \centering
\includegraphics[width=0.8\linewidth]{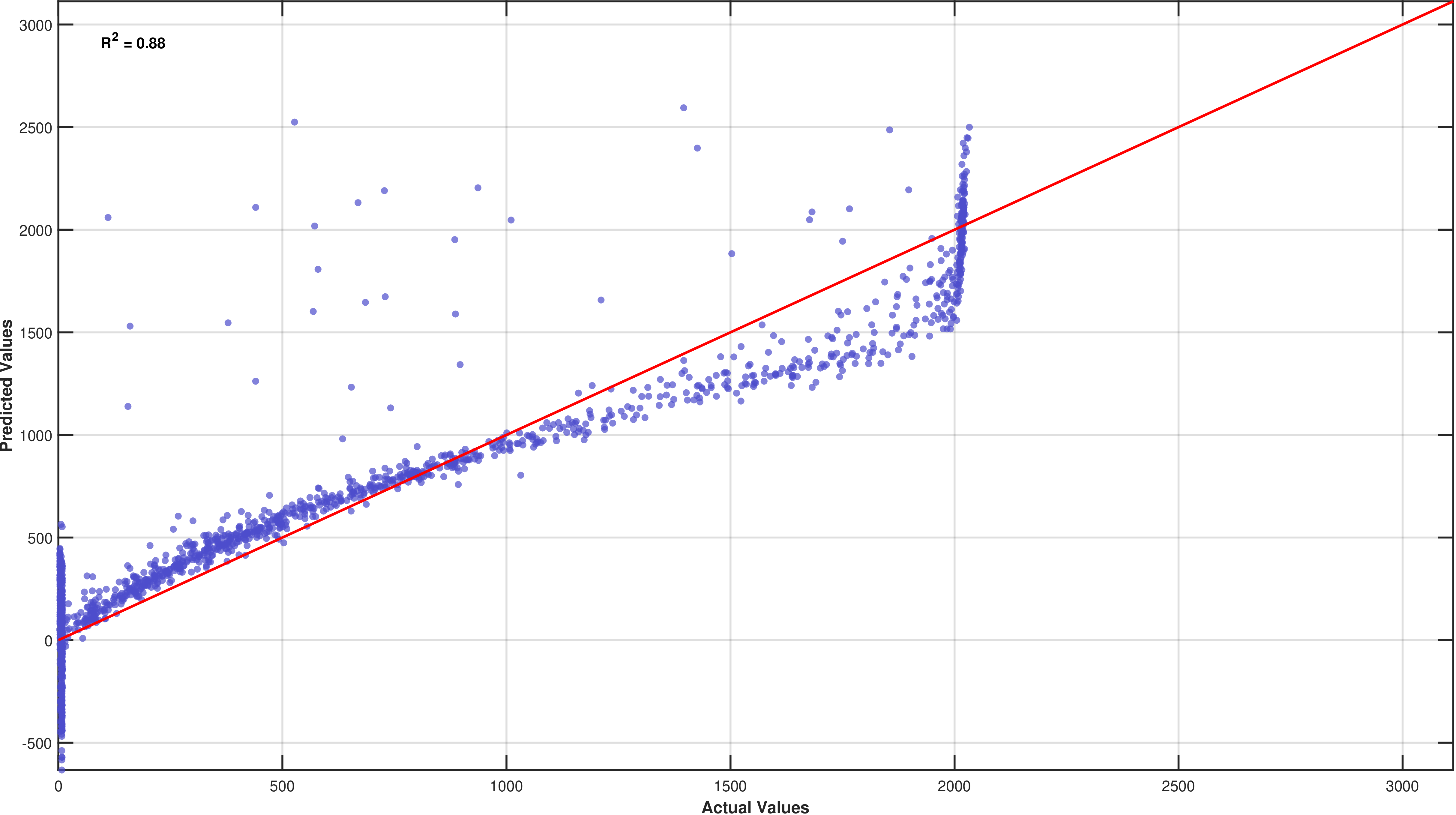}
    \caption{Linear Regression}
    \label{fig:cls_linear_pred}
\end{subfigure}

\caption{Prediction performance of QNN and classical ML models.}
\label{fig:act_vs_pred_fig}
\end{figure}

\end{samepage}

\clearpage
\twocolumn

\end{document}